\documentstyle[sprocl,epsf]{article} 

\def\be{\begin{equation}} 
\def\ee{\end{equation}} 
\def\bea{\begin{eqnarray}} 
\def\eea{\end{eqnarray}} 

\newcommand{\bef}{\begin{figure}} 
\newcommand{\eef}{\end{figure}} 
\newcommand{\hmp}{ h^{-1}Mpc} 
\newcommand{\etal}{et al.} 
\def\spose#1{\hbox to 0pt{#1\hss}} 
\def\ltapprox{\mathrel{\spose{\lower 3pt\hbox{$\mathchar"218$}} 
 \raise 2.0pt\hbox{$\mathchar"13C$}}} 
\def\gtapprox{\mathrel{\spose{\lower 3pt\hbox{$\mathchar"218$}} 
 \raise 2.0pt\hbox{$\mathchar"13E$}}} 
\def\inapprox{\mathrel{\spose{\lower 3pt\hbox{$\mathchar"218$}} 
 \raise 2.0pt\hbox{$\mathchar"232$}}} 
 
\begin{document} 
 
\title{Reconstructing the cosmological puzzle 
 \footnote{ In the proceedings of the  
"VI Colloque de Cosmologie" 
Paris 16-18 June 1999}} 
 
\author{Luciano Pietronero$^1$ and Francesco Sylos Labini$^{2,1}$} 
 
\address{      	 
		$^1$INFM Sezione Roma1,         
		Dip. di Fisica, Universit\'a "La Sapienza",  
		P.le A. Moro, 2,   
        	I-00185 Roma, Italy.  
        	\\ 
        	$^2$D\'ept.~de Physique Th\'eorique,  
		Universit\'e de Gen\`eve,   
		24, Quai E. Ansermet, CH-1211 Gen\`eve, Switzerland. 
		}

\maketitle 
\abstracts{ In the debate about  galaxy correlation 
there are different questions which can be 
addressed separately: Which are the statistical 
methods able to properly detect scale invariance  
and describe, in general, the properties 
of irregular and regular distributions ? 
Which are the implications for cosmology 
of the fractal behavior of galactic structures,
 up to a certain scale $\lambda_0$ ? 
Which is the homogeneity scale $\lambda_0$, i.e. the scale beyond 
which galaxy distribution has an eventual crossover to homogeneity ?  
These are three different, but related, problems, 
which must be considered in different steps,   
from the point of view of data analysis as well as 
from the  theoretical perspective.  
}

\section{Introduction}

Nowadays there is a general agreement about the fact 
that galactic structures are fractal up to a distance 
scale of $\sim 30 \div 40 \hmp$ \cite{slmp98,jmsl99} and the  
increasing interest about the fractal versus  
homogeneous distribution of galaxy in the last year 
 \cite{coles98,scara98,rees99,cappi98,martinez99,hutton99,chown99,landy99} 
has focused, mainly on the determination of the homogeneity 
scale $\lambda_0$.\footnote{See the web page {\it  
http://pil.phys.uniroma1.it/debate.html } where  
all these materials have been collected}
Instead, we would like to discuss three 
important and different aspects of this problem 
which, we believe,  have not  been considered  appropriately  
in the debate. The main point we would like to stress
is that galaxy structures are fractal no matter what is
the crossover scale, and this fact has never been
properly appreciated.
 
\begin{itemize} 
 
\item {\it Methodological point.} 
 
The major problem from the point of view 
of data analysis is to use statistical methods 
which are able to properly characterize scale 
invariant distributions, and hence which are 
also suitable to characterize an eventual 
crossover to homogeneity.  
Our main contribution \cite{pie87,cp92,slmp98}, 
in this respect, has been to clarify that the usual  
statistical methods (correlation function, 
power spectrum, etc.) are based on the assumption of  
homogeneity and hence are not appropriate 
to test it. Instead, we have introduced and developed 
various statistical tools which are able 
to test whether a distribution is  
homogeneous or fractal, and to correctly characterize 
the scale-invariant properties. 
Such a discussion is clearly relevant also 
for the interpretation of the properties 
of artificial simulations. The agreement about  
the methods to be used for the analysis of future 
surveys such as the Sloan Digital Sky Survey (SDSS) 
and the two degrees Fields (2dF) is clearly a fundamental  
issue. 
 
\item {\it Implication of the fractal structure up 
to scale $\lambda_0$}. 
 
 The fact that galactic structures 
are fractal, no matter what is the homogeneity scale $\lambda_0$, 
 has deep implication on the interpretation 
of several phenomena such  
as the luminosity bias, the mismatch 
galaxy-cluster, the determination of the average  
density, the separation of linear and  
non-linear scales, etc.,  
and on the theoretical concepts used 
to study such properties. For example the properties 
of dark matter are inferred from the ones of visible 
matter, and hence they are closely related. 
If now one observes different  
statistical properties for galaxies and clusters, 
this necessarily implies a change of perspective 
on the properties of dark matter. 
 
\item {\it Determination 
of the homogeneity scale $\lambda_0$. } 
 
This is, clearly, a very important point  
which is at the basis of the understanding 
of galaxy structures and more generally 
of the cosmological problem. We distinguish 
here two different approaches: direct tests and  
indirect tests. By direct tests, we mean the determination 
of the conditional average density in three dimensional surveys, 
while with indirect tests we refer to other possible analyses, 
such as the interpretation of angular surveys, the  
number counts as a function of magnitude or of distance  or, 
in general, the  
study of non-average quantities, i.e. when the fractal
dimension is estimated without making an average 
over different observes (or volumes).
While in the first case one is able to have a  
clear and unambiguous answer from the data, in the second 
one is only able to make some weaker  
claims about the compatibility of the data 
with a fractal or a homogeneous distribution. 
However, also in this second case, it is possible to understand  
some important properties of the data, and to 
clarify the role and the limits of some underlying  
assumptions which are often used without a  
critical perspective. 
  
\end{itemize}

\section{Statistical Methods} 
 
 The proper methods to characterize irregular as  
 well as regular distributions have been 
 discussed    in Coleman \& Pietronero \cite{cp92} 
 and Sylos Labini \etal \cite{slmp98}  
 in a detailed and exhaustive way. 
 The basic point is that, as far as a system  
 shows power law correlations, the usual $\xi(r)$ analysis  
\cite{pee80} gives an incorrect result, since it is  
 based on the a-priori assumption of homogeneity.  
 In order to check whether homogeneity is present  
 in a given sample one has to use  
 the conditional density $\Gamma(r)$  
 defined as \cite{pie87} 
 \be 
 \label{eq1} 
 \Gamma(r) = \frac{\langle n(r_*) n(r_*+r) \rangle}{\langle n  
 \rangle} = 
 \frac{BD}{4 \pi} r^{D-3} 
 \ee 
 where the last equality holds in the case of a fractal 
 distribution with dimension $D$ and pre-factor $B$.  
 In the case of an  
 homogenous distribution ($D=3$) the conditional density  
 equals the average density in the sample.  
 Hence the conditional density is the suitable  
 statistical tool to identify  
 fractal properties (i.e. power law correlations 
 with codimension $\gamma=3-D$) as well as  
 homogeneous ones (constant density with sample size). 
 If there exists a transition scale  $\lambda_0$  
 towards homogenization, 
 we should find $\Gamma(r)$ constant for scales  
 $r \gg\lambda_0$. 
 
\subsection{Other (indirect) methods to detect homogeneity} 
 
Basically $\lambda_0$ is related to the maximum 
size of voids: the average density will be constant, 
at least, on scales larger than the maximum 
void in a given sample. Several authors have  
approached this problem by looking at  
voids distribution. For example 
El-Ad and Piran (1997) have shown that the SSRS2  
and IRAS 1.2 Jy. redshift surveys are dominated  
by voids: they cover the $\sim 50 \%$ of the volume. 
Moreover the two samples show very similar  
properties even if the IRAS voids 
are $\sim 33 \%$ larger than SSRS2 ones because they  
are not bounded by narrow angular limits as the SSRS2 voids.  
The voids have a scale of at least $\sim 40 \div 50 \hmp$ 
and the largest void in the SSRS2 sample has a 
diameter of $\sim 60 \hmp$, i.e. comparable 
to the Bootes void. The problem is to understand whether 
such a scale has been fixed by the samples' volume, 
or whether there is a tendency  not to find larger 
voids: in this case one would have a (weaker evidence) for the  
homogeneity scale. In any case, we note that 
the homogeneity scale cannot be smaller 
than the scale of the largest void found in these samples 
and that one has to be very careful when comparing the size of 
the voids to the effective depth of catalogs. For example  
in the Las Campanas Redshift Survey, even if  
it is possible to extract sub-samples limited at $\sim 500 \hmp$, 
 the volume of space investigated is not so 
large, as the survey is made by thin slices. In  
such a situation a definitive answer to the dimension 
of the of voids, and hence to the existence of the homogeneity 
scale, is rather difficult and uncertain.

Another complementary way to study the  
eventual crossover to homogeneity  of galaxy distribution  
is represented by the morphological signatures identified  
by tools such as the Minkowski Fuctionals. 
Kerscher et al. (1998), by analyzing the IRAS samples 
have found that there are  large 
fluctuations in the clustering  
properties as seen in a large  
difference between the northern and southern parts of 
the catalogue on scales of $\sim 100h^{-1}Mpc$.  
These fluctuations remain discernible even on the scale of 
$200h^{-1}Mpc$ and this 
is again a sign of the inhomogeneous character of galaxy 
structures at these scales. There are several other approaches to 
this problem, but we believe that the analysis via 
the conditional average density is the more stable 
and powerful to understand the correlation and statistical 
properties of  
a given sample of galaxies.

\subsection{The standard correlation function} 
 
 It is simple to show that  
 in the case of a fractal distribution the  
 usual $\xi(r)$ function in a spherical sample of radius $R_s$  
 is \cite{pie87,cp92} 
 \be 
 \label{eq2} 
 \xi(r) = \frac{D}{3} \left(  
 \frac{r}{R_s} \right)^{D-3} -1 \; . 
 \ee 
 From Eq.\ref{eq2} we can see two main problems of the  
 $\xi(r)$ function: its amplitude depends on the sample size $R_s$ 
 (and the so-called correlation length $r_0$, defined as  
 $\xi(r_0) \equiv 1$, linearly depends on $R_s$)  
 and it {\it has not} a power law behavior. 
 Rather the power law behavior is present  
 only at scales $r  \ll r_0$, and  
 then it is followed by a sharp break in the log-log plot 
 as soon as $\xi(r) \ltapprox 1$. Such a behavior does not 
 correspond  to any real change of the correlation  
 properties of the  
 system (that is scale-invariant by definition),
i.e the break is artificial,  and it  
 makes extremely difficult the estimation 
 of the correct fractal dimension. 
 In particular if the sample size is not large enough  
 with respect to the  
 actual value of $r_0$,  
 the codimension estimated by the $\xi(r)$ function  
 ($\gamma \approx 1.7$)  
 is systematically larger than $3-D$ ($\gamma \approx 1$)  
 \cite{slmp98}.

 Given this situation it is clear that the $\xi(r)$ analysis  
 is not suitable 
 to be applied unless 
 a clear crossover  towards homogenization  
 is present in the samples analyzed. As this is not the case,  
 it is appropriate and convenient to use  
 $\Gamma(r)$ instead of $\xi(r)$. We have discussed  
 in detail  that the use of the  
 correct statistical metho{ds \cite{slmp98}  
 is complementary  
 to a change of perspective from a theoretical 
 point of view.  
 
\subsection{Properties and limits of real catalogs} 
 
Before we discuss some determination 
of the conditional average density in real surveys, 
we briefly recall the properties of three dimensional data. 
A catalog is usually obtained by measuring the redshifts  
of all the galaxies with apparent magnitude brighter 
than a certain apparent magnitude limit $m_{lim}$, 
 in a certain region of the sky defined by a solid angle $\Omega$. 
An important selection effect exists, in that at every distance in the  
apparent magnitude limited survey, there is a definite limit in intrinsic  
luminosity 
which is the absolute magnitude of the fainter galaxy 
which can be seen at that distance. Hence at large distances, intrinsically 
faint objects are not observed whereas at smaller distances they are observed. 
In order to analyze the statistical properties of galaxy distribution, 
a catalog which does not suffer for  
this selection effect 
must be used. In general, it exists a very well known procedure to 
obtain a sample that is not biased by this luminosity selection effect: 
this is the so-called {\it  "volume limited"} (VL) sample. 
A  VL  sample contains every galaxy in the volume 
which is more luminous than a certain limit, so that in such a 
 sample 
there is no incompleteness for an observational 
luminosity selection effect \cite{dp83,cp92}. 
Such a sample is defined by a certain maximum distance $R_{VL}$ 
and the absolute magnitude limit $M_{VL}$  
given by 
\be 
\label{e45} 
M_{VL}=m_{lim}-5\log_{10}R_{VL} -25 -A(z) 
\ee 
where $A(z)$ takes into account various corrections (K-corrections,  
absorption, relativistic effects, etc.), and  
$m_{lim}$ is the survey apparent magnitude limit. 
 
In a give sample $\Gamma(r)$ can be computed 
in a range of scale defined by a lower and an upper 
cut-off, which are defined in the following way. 
 
(i) The upper cut-off $R_s$ up to which 
the statistic can be calculated. It is simply 
the size of the largest sphere around any galaxy  
which can be inscribed inside the sample volume,  
since the average conditional density is computed  
in complete shells\cite{cp92,slmp98}. 
 It clearly depends on the  
survey geometry - on the solid angle of the survey  
and the effective depth of the particular sub-sample 
we analyze. Note that as we approach this upper cut-off 
the number of independent spheres being averaged over  
decreases rapidly. This leads to a systematic error 
at $r \sim R_s$  (which depends on the unknown  
underlying fluctuations in the quantity being averaged) 
which is difficult to quantify \cite{jmsl99}.   
 
(ii) A lower cut-off $\langle \Lambda \rangle$, 
which is related to the number of points contained  
in the sample. It is simply the scale below which 
the behavior of the conditional density is  
dominated by the sparseness of the points. 
Since there are typically no points  
at sufficiently small distances in the neighborhood  
of any given one, we expect 
$\Gamma(r)$ to fluctuate back and forth to  
zero, and $\Gamma^*(r)$ to decay away from any finite  
value as the volume $1/r^3$.  
A definition of this scale \cite{slmp98}, 
appropriate both to the case of fractal structures and 
homogeneous ones, is the average  
distance between nearest neighbors.  
 
\subsection{Results} 
 
In Fig.1 we show the results of the  
analysis of all the available galaxy samples 
through the conditional density \cite{slmp98,jmsl99,jmslp99}, 
\bef 
\epsfxsize 8cm  
\epsfysize 8cm  
\centerline{\epsfbox{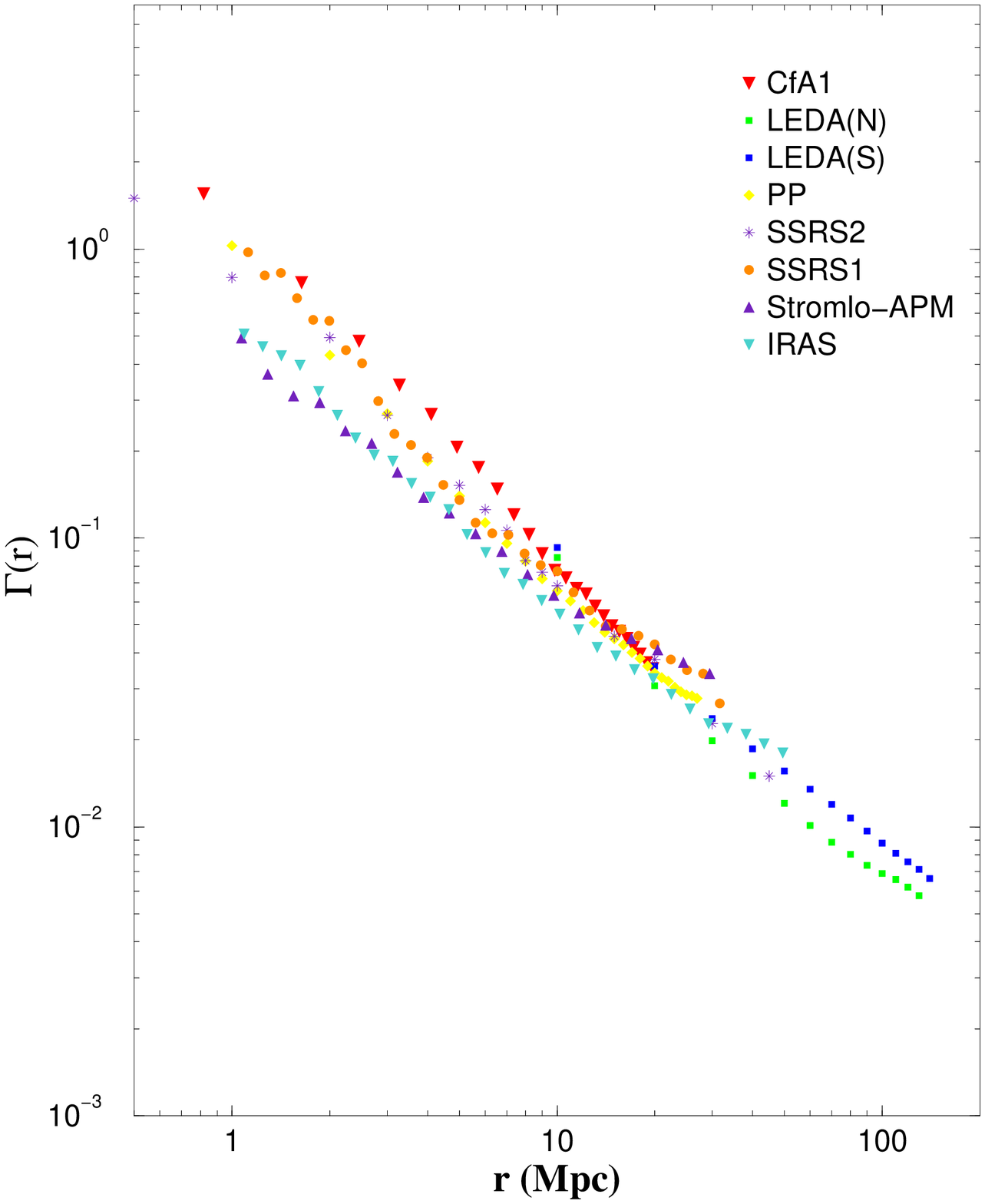}}  
\caption{  
Conditional average density computed for various 
different galaxy surveys (from Sylos Labini et al. (1998)). 
The power law behavior corresponds to a fractal structure with dimension 
$D \approx 2$} 
\eef 
\bef 
\epsfxsize 8cm  
\epsfysize 8cm  
\centerline{\epsfbox{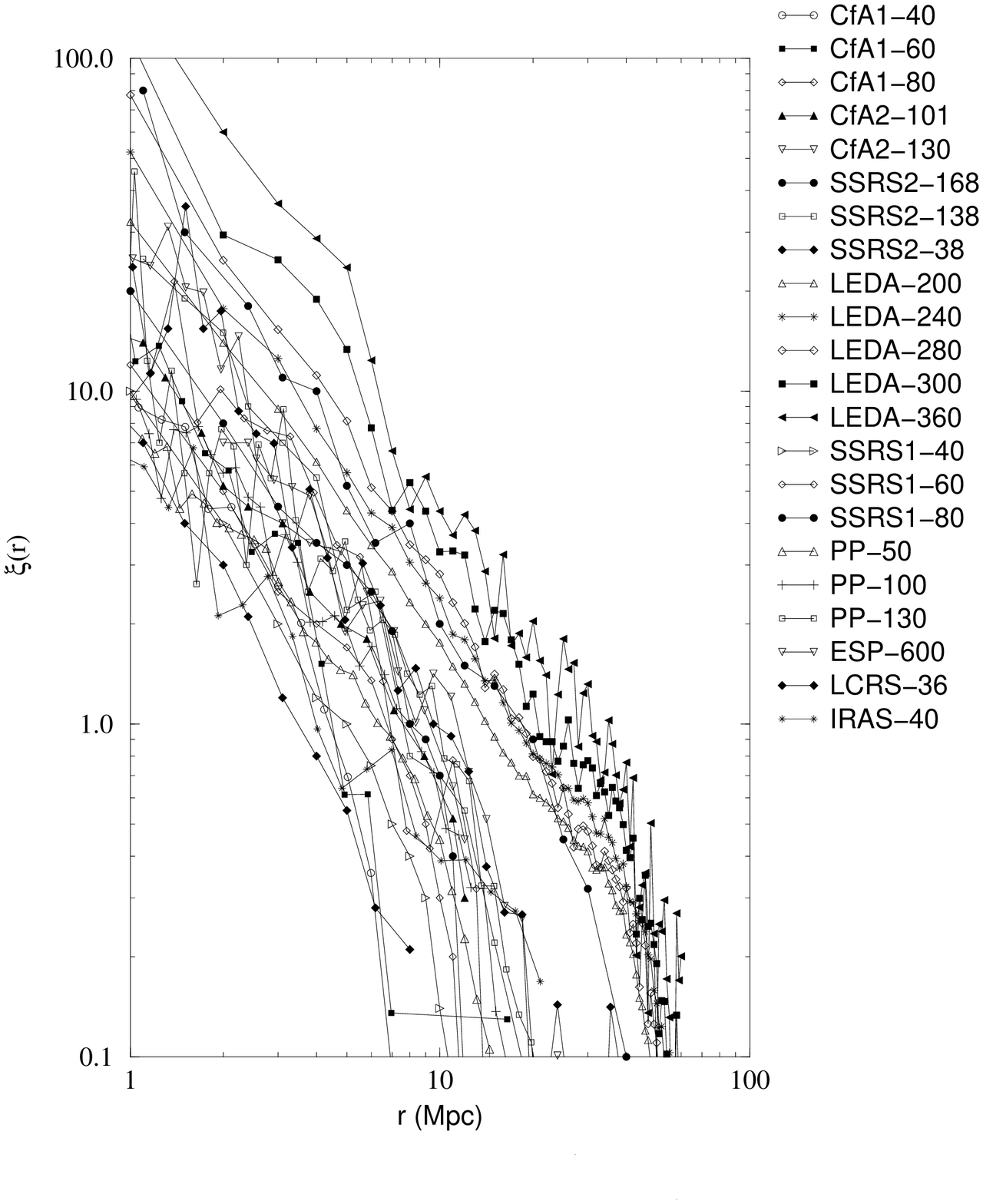}}  
\caption{  
The standard correlation function $\xi(r)$,  
computed for the same galaxy samples of Fig.1  
(from Sylos Labini et al. (1998)).} 
\eef 
while in Fig.2 we show the behavior of the standard $\xi(r)$ 
in the same catalogs. One may note that the different data are in rather good 
agreement when analyzed by $\Gamma(r)$ and give  
a complex information when seen from the perspective of $\xi(r)$. 
As we discuss below, this complex situation has given rise to some 
confused concepts as the luminosity bias or 
mismatch galaxy-cluster. 
 
\subsection{Non-average quantities} 
 
Another possible way of measuring the fractal dimension can be done by using 
the mass-length 
 relation between $N(<R)$, the number of points inside a portion of sphere of  
 radius $R$ with solid angle $\Omega$, and the distance $R$,
which can be written as 
\be  
\label{new1a}  
\langle N(<R) \rangle  = B R^D \frac{\Omega}{4\pi} \;. 
\ee  
Eq.\ref{new1a} holds for {\it average } quantity, while we would 
like to understand which kind of fluctuations 
affect such a behavior in the case we do not 
perform an average.  
We can identify two basic kinds of fluctuations: 
the first ones are intrinsic $f(R)$  and are due to 
the highly fluctuating nature of fractal distributions 
while the second ones are Poissonian fluctuations which we do not 
consider here (see Sylos Labini et al., (1998) for a more detailed 
discussion). 
Concerning the first ones,  
one has to consider that  
the mass-length relation is a convolution of fluctuations 
which are present at all scales. 
For example one encounters, at any scale,  a  
large scale structure  and  
then a huge void: these fluctuations   affect the power law behavior of  
$N(<R)$. We can quantify these effects as a {\it modulating term}  
around the expected average given by Eq.\ref{new1a}.  
Therefore, in the observations from a single point {\it "i"}  
we   have  
\be 
\label{cazzz4}  
N(<R)_i = B R^D \frac{\Omega}{4\pi} \cdot f_{\Omega}(R, \delta \Omega) \; . 
\ee  
This equation  shows that the amplitude of $N(<r)_i$ is related to  
the amplitude of the intrinsic fluctuations and not only  
to the lower cut-off $B$. 
In general this 
fluctuating term  depends on the direction of observation $\Omega$  
and on  the solid angle of the survey $\delta \Omega$ 
so that $f(R) = f_{\Omega} (R, \delta \Omega)$. 
If we perform the ensemble average (i.e. over non overlapping volumes)  
of this fluctuating term we  
can smooth out its effects: In such a way 
 the conditional density, averaged over all the points 
 of the sample,  has a single power law behavior.  
From the above discussion it is clear that the determination 
of $N(<R)_i$ is much more problematic, i.e. subjected 
to fluctuations, than the full average, and hence its 
information is much more weaker from a statistical point of view. 
 
There are several other methods and statistical tools as for example 
the angular correlation function, the three dimensional 
power spectrum, and we refer the interested reader to  
Sylos Labini et al. (1998) for a more exhaustive discussion 
of this matter.


\section{Implications for the statistical methods and theoretical concepts}

We now consider some specific points  
which are discussed in the papers  
\cite{coles98,scara98,rees99,cappi98,martinez99,hutton99,landy99} 
and which can be interpreted in a more general perspective. 
 
\subsection{Cosmological Principle} 
There is common  
confusion about the 
Cosmological Principle(CP)\cite{rees99}: That homogeneity is necessary to 
satisfy it, and in particular that a fractal distribution  
contradicts it. Understood as a principle which states the 
equivalence of all points, the CP only implies homogeneity 
when one assumes analyticity.  
 
More specifically, 
it is quite reasonable to assume that the earth  
is not at a privileged position 
in  the universe and to consider this as a principle, the CP. The  
usual   implication of this principle is that the universe  
must be homogeneous. This reasoning implies the hidden assumption  
of analyticity that often is not even mentioned. In fact, the above  
reasonable requirement only leads to {\it local isotropy}. 
For an analytical  
structure this also implies homogeneity. However, if  
the structure is  
not analytical, the above argument  does not hold. For example, a  
fractal structure is locally isotropic but not homogeneous. 
This means  
that a fractal structure satisfies the CP in the sense that all the points  
are essentially equivalent (no center or special points), but this does  
not imply that these points are distributed uniformly 
\cite{man77,cp92,slmp98}. 
 
Einstein's equations can be solved by assuming 
a constant density  and the well-known Friedmann 
solutions are in fact the simplest ones. 
However this does not imply that one could not 
find  different solutions of the field's equations. 
 One way to still obtain solutions to Einstein's equations is 
to assume that the inhomogeneity is simply a small "perturbation" 
on a homogeneous Universe. However, a fractal Universe is more 
than a mere perturbation--it is a radically different kind of 
Universe. This opens a new perspective and 
one should focus  
the theoretical investigation  
on perturbed solutions and average quantities \cite{buchert,jampsl99}.

\subsection{Angular data} 
 
All the large scale structures  
of galaxies and galaxy clusters  
 have been discovered by redshift data. Never these findings were 
found from the angular data. 
The angular data are intrinsically incomplete, it is not a matter of 
statistical fluctuations (for which having more data is of help). 
If it would be possible to get the three dimensional  
information  from the angular the measurement 
of redshifts would be useless. 
For instance, from the shadow 
of clouds it is not possible to derive the fractal dimension of the 
cloud (3-d property), no matter how many data one has. 
If one accepts the idea that from projections one can reconstruct the real 
data one can go on to the paradox of projecting on a line (instead of a 2-d 
plane) and then even on a single point. So there must be intrinsic 
limitations to this process that has nothing to do with the statistical 
validity of the data. This intrinsic limitations are qualitative and not 
statistical and they have never been really addressed with specific tests. 
 
In the analysis of the angular data, without a measured 
space coordinate, the inference of three dimensional properties 
(such as homogeneity) relies on a deconvolution which is only 
possible with the assumption of large scale homogeneity.  
In particular the strong constraint of the  APM survey 
on fluctuations at $100 h^{-1}$ Mpc\cite{rees99} 
comes from an angular survey.  
In this respect  
we have clarified in various papers \cite{cp92,slmp98,deslmp97}  
the angular properties of a fractal and in others  
the correlation properties  
of two subsamples of the APM catalog (which are  
the only published data: APM Bright Galaxies \cite{msl97}, 
which has only the angular coordinates, 
and APM-Stromlo \cite{slm98} which has the redshifts) 
 
In summary, we would like to stress  that: 
 
(i) The fact that galaxy structures and voids have been  
discovered by redshift measurements  
represents an important point in the interpretation 
of angular catalogs. 
The angular projections are too smooth  
and galaxy structures only appear in the  
three dimensional catalogs.

(ii) If one would have been able to reconstruct  
the three dimensional statistical properties  
from the angular ones {\it without any assumptions} 
then the measurements of redshift would have  
been useless, or at least they would have only 
reduced the error bars in the estimation 
of the correlation function. 
This is clearly not the case 
and in all the "reconstruction" of 3-D properties 
from angular ones one is forced to make some 
{\it untested assumptions}. For example the famous rescaling 
of the amplitude of the angular correlation function in the APM catalog 
\cite{pee93} is a typical result whose interpretation 
depends on some hidden  assumptions. 
The angular correlation function is in fact a convolution 
of the three point correlation function in the three 
dimensional space and moreover one is not performing 
an average over different observers but only 
over different pairs at a certain angular separation. 
Hence the amplitude of the angular correlation 
function, being not average out 
over different observers, 
is strongly affected by the intrinsic 
fluctuations of the fractal structure. 
The difference between average and non-average  
quantities for a fractal has never been appreciated 
in this respect \cite{slmp98} (see below). 
 
(iii) We would like to stress that the APM angular catalog 
is  still not available, after ten years from the publication 
of the data analysis \cite{maddox90}. However, the ``reconstructed'' 
three dimensional correlation function from the  
two-dimensional one computed in the APM survey, 
is still believed to be one of the best estimate 
of the correlation properties of galaxies \cite{jenkins98}. 
The underlying  
idea of these analyses is that photometric catalogs  contain 
many more galaxies than do redshift surveys. 
This advantage in statistics is often considered  
enough to offset the extra  
information about distance contained in redshift surveys \cite{dodelson99}. 
In Montuori \& Sylos Labini  \cite{msl97} we have demonstrated that 
$\omega(\theta)$ suffers from the same biases of $\xi(r)$, 
and that the information about correlations it gives 
is incorrect as it is the one three dimensional $\xi(r)$. 
We have also shown \cite{slm98}  that the APM-Stromlo 
redshift surveys, which is a subsample  
extracted from the angular APM, does not 
show any intrinsic characteristic 
length: instead it presents power law correlation up to 
$\sim 40 \hmp$ with fractal dimension $D \approx 2$.  
This result has been confirmed 
by Hatton\cite{hutton99}, even if he has concluded that 
a tendency to homogeneity is detected at scales  
larger than $\sim 40 \hmp$ (i.e. he found  
an increase in the fractal dimension in the sparser 
sample which contradicts the estimations 
performed, at the same scales, in the samples 
with a larger number of galaxies). Even if such a situation 
would be true, and then the fractal dimension would 
approach to $3$ at scales $r > 40 \hmp$, this would imply that  
the ``standard'' results of $r_0 = 5 \hmp$ 
or $D=1.2$ at scales $0.1-10 \hmp$, or  the  
whole reconstruction of the 3-d from the 2-d 
properties, are incorrect and based on  
assumptions which are not verified by 
a more appropriate test.

\subsection{Average and non-average quantities} 
 
As we have discussed in Sec.2, in the discussion of  
three dimensional data  one must take into account that 
the  definition of the fractal dimension $D$,   
 is the  one  referring to averaged quantities. 
The constraint on $D$  
at scales larger than $\sim 100 h^{-1}$ Mpc  \cite{scara98,rees99} 
comes from a number count from the origin.

For example, in the paper of Scaramella et al.\cite{scara98} 
there is the claim that galaxy distribution is homogeneous  
at scales larger than $\sim 300 \hmp$. 
However, the results for the number 
count dimension in this paper are in fact highly varying  
in the various samples, ranging from 2.5 to 3.5. Further the 
deeper samples omitted in the paper of Scaramella et al.\cite{scara98} 
in fact show 
a dimension of $4$ or even more. This behaviour is in fact  
consistent with the fluctuations characteristic of a fractal, 
which have not be averaged out in the number count. These issues  
are discussed in a recent paper by our group \cite{jmslp99}. 
 
Actually the main point is that, if  
such a cross-over exists as described by the authors \cite{scara98}, 
the scale characterizing it is $\sim 100 \div 300 \hmp$. 
This  invalidates the ``standard''  
analysis of the same  
catalogue given elsewhere  
by the ESP collaboration \cite{guzzo99} which results in a ``correlation 
length'' of only $r_0 = 4 \hmp$ (see Fig.3).  
\bef 
\epsfxsize 10cm  
\epsfysize 10cm  
\centerline{\epsfbox{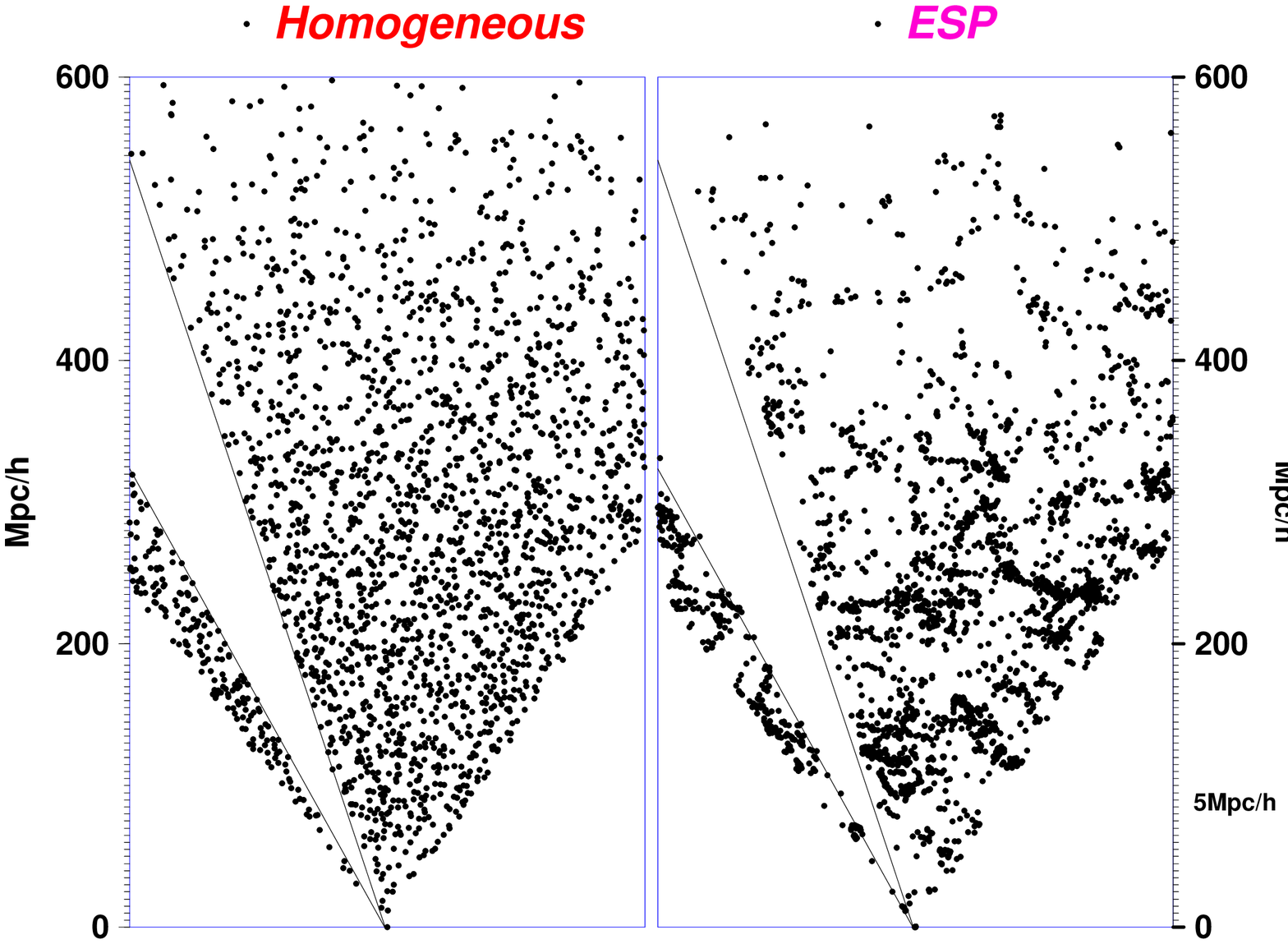}}  
\caption{Right Panel.  
Redshift space distribution of galaxies 
ESP survey (Vettolani et al., 1997). 
The total solid angle of the survey is $\Omega \sim 0.006 sr$ 
and the apparent magnitude limit is $m_{B} = 19.4$.  
This strip is $1^{\circ} \cdot 20^{\circ}$ tick. 
In this regaion there are 3175 galaxies. 
The conical empty region on the left is  
due to an observational effect. 
Left Panel. Redshift space distribution of galaxies of an homogeneous 
distribution of galaxies with the same selection effects 
of the ESP survey (Right Panel). The homogeneity scale 
is about $\sim 10 \hmp$ 
} 
\eef 
Furthermore we have shown \cite{jmslp99} that the evidences for 
a cross-over to homogeneity rely on the choice of 
cosmological model, and most crucially on the 
so called K corrections.  
In particular we have demonstrated that 
 the $D \approx 3$ 
behaviour seen in the K-corrected data 
of Scaramella et al.\cite{scara98} is in fact 
unstable, increasing systematically towards $D=4$ 
as a function of the absolute  
magnitude limit. This behaviour 
can be quantitatively explained as the effect of  
an unphysical 
K-correction in the relevant range of red-shift ($z \sim 0.1 
\div 0.3$). A more consistent interpretation of 
the number counts is that $D$ is in the range $2 \div 2.5$, 
depending on the cosmological model, consistent with  
the continuation of the fractal $D\approx 2$ behaviour 
observed at scales up to $\sim 100 \hmp$. 
This implies a smaller K-correction. 
In this case the detection of the fractal 
behavior relies on the determination of non-average 
quantities and hence, from a statistical point of view, 
this is much weaker than the full three dimensional  
analysis. The only way to improve such determinations 
is to have larger catalogs of galaxies (SDSS and 2dF). 
 
The same comment about non-averaged quantities holds  
for the results 
which are considered to show  
an {\it increase of fractal dimension with scale}.  
 These are all 
behaviors observed at length scales where one cannot average over  
a number of independent points in the samples, and are best  
interpreted as the effect of non-averaged fluctuations  
(note the large variation). From these measurements alone 
is very difficult to conclude whether the fractal dimension 
is really increasing or if this effect is just due to the  
effect of intrinsic fluctuations. 
 
For instance, Cappi \etal \cite{cappi98} 
by analyzing the SSRS2 catalog 
have measured that $\Gamma(r)$ has a power 
law behavior up to $\sim 40 \hmp$
and they have cocnluded that the samples
apporaches homogeneity at larger scales.
Let's see in more detail this claim \cite{mjsl99}. 
The SSRS2 redshift survey is an important new 
probe of the local universe. The catalog consists  
of $\sim 3500 $ galaxies, is complete up to  
$m_B = 15.5$, and covers a solid angle of  
$\Omega \approx 1.13 \; sr$. 
Using the conditional density we have shown 
\cite{mjsl99}  that the   
galaxy distribution in the sample 
has well defined scale invariant properties 
in the range ($\sim 1 \div 40 \hmp$) in which  
it can be analyzed with this statistic. The corresponding  
fractal dimension is found to be $D = 2.0 \pm 0.1$, in good  
agreement with various other galaxy catalogs analyzed using 
the same methods, in agreement with the results of Cappi et al\cite{cappi98}.
No evidence for a characteristic scale  
for galaxy clustering is found up to $\sim 40 \hmp$. 
At larger scales the  
number counts from the origin are analyzed, and  
typically larger but highly fluctuating dimensions  
are found ($D \sim 2 \div 4$). We have provided evidences 
that these  
are better interpreted as the fluctuating behavior  
characteristic of a continuing fractal structure rather  
than as an indication of a cross-over to homogeneity.  
 
Then, as previously mentioned, if $\Gamma(r)$ has a power law 
behavior, $r_0$ is not a characteristic length, 
and hence studying the eventual behavior 
of $r_0$ for galaxies of different 
luminosity is a completely misleading and  
incorrect analysis (see below). 
 
\subsection{Luminosity Bias} 
 
We would like to stress again that, even if the  
fractal behavior breaks at a certain scale $\lambda_0$, 
the use of $\xi(r)$ is in any inconsistent at scales 
smaller than $\lambda_0$.  
 
One of consequence, which have never been appreciated of 
the fractal behavior of galaxy distribution is the following: 
as long as $\Gamma(r)$ shows a power law behavior, then 
the use of $\xi(r)$, or its power spectrum, is completely 
misleading. All the properties inferred by the $\xi(r)$ 
analysis are artifacts. For example, the fractal dimension 
estimated by the log-log plot of $\xi(r)$ is systematically 
smaller than the (correct) one found by looking at $\Gamma(r)$ 
\cite{slmp98}. Also all the characteristic scales 
associated to $\xi(r)$ are just fraction of the sample size.

For example, the fact that Cappi et al.\cite{cappi98}  
have detected a behavior 
of $r_0$ with sample size that is not in agreement with  
the linear scaling of a fractal is probably due to the following reason 
(see Montuori et al. \cite{mjsl99} for a more detailed discussion). 
When one normalizes the conditional average density to the mean  
density in the sample in order to compute $\xi(r)$, one  
is performing a very delicate operation from which 
it depends the amplitude of $\xi(r)$ itself.  
In fact, the average number density of galaxies is just given by the  
total number divided by the volume of the sample. 
However,  
if the distribution is fractal, even with a cut-off 
to homogeneity at a scale comparable with the size of the  
sample itself, than this number, which is not average out, 
can have a fluctuations of order one, due to the 
intrinsic fluctuations of the fractal \cite{slmp98,mjsl99}, 
making the estimation of the amplitude of $\xi(r)$  
completely useless. If, and only if, $\Gamma(r)$ has a 
clear cut-off to homogeneity (for a decade or so) 
then one may use $\xi(r)$ to study the correlation 
properties of the fluctuations from the average density! 
 
\subsection{Power Spectrum of density fluctuations} 
 
The problems  
with the standard correlation analysis also show that the  
properties of fractal correlations have not been really appreciated. These  
problems are actually far more serious and fundamental than mentioned, 
for example, by  
Landy \cite{landy99}  
and the idea that they can be solved by simply taking the Fourier  
transform is once more a proof of the superficiality of the discussion.  
We have extensively shown \cite{sla96,slmp98} that 
the power spectrum of the density fluctuations 
has the same kind of problems which $\xi(r)$ has, 
because it is normalized to the average density as well. 
The density contrast $\delta(r)=\delta \rho(r)/ \langle \rho \rangle$ 
is not a physical quantity unless the average density 
is demonstrated to exist.  
More specifically, like in the case of $\xi(r)$, the power spectrum 
(Fourier Transform of the correlation function) 
is affected by finite size effects at large scale:  
even for a fractal distribution the power spectrum 
has not a power law behavior but it shows 
a large scale (small $k$) cut-off which  
is due to the finiteness of the sample \cite{sla96}. 
Hence the eventual detection of the turnover of the power 
spectrum, which is expected in CDM-like models  
to match the galaxy clustering to the anisotropies 
of the CMBR, must be  considered a finite size effect, 
unless a clear determination of the average density 
in the same sample has been done.

\subsection{CMBR anisotropies} 
 
A typical objection to our result  
concerns the compatibility of the fractal with the 
highly isotropic CMBR \cite{rees99}. 
We point out  that  
the isotropy of the COBE data refers to the background 
radiation: the relation between this radiation and the matter distribution, 
corresponding to the inhomogeneous properties of galaxy clustering, 
requires a complex theory with many assumptions. Considering that matter 
and radiation have a completely different origin their relations should be 
considered with great caution. In particular if the present theory cannot 
explain these two observations one should try to improve the theory instead 
of dismissing one of the observations. A new perspective 
on this problem has been recently addressed by our group \cite{jampsl99}.

\section{Theoretical implications: correlation and bias} 
 
We have discussed the concept of correlation and bias, 
as it usually defined in the literature, 
in a series of papers\cite{gsld99,gsl99}. 
We review here the main points of this discussion.
The concept of bias, i.e. the relative abundance  
and distribution of objects  
of different mass, has been  originally  
introduced by Kaiser (1984) to explain the different 
amplitudes of the correlation function $\xi(r)$ 
found for galaxies and galaxy clusters. 
Afterwards it has also  been   invoked to explain the increasing 
amplitudes of $\xi(r)$ 
for galaxies with brighter luminosity. Finally, it is   
used to describe the ``clustering'' of  dark matter  relative 
to the one of visible matter. In general it is believed that objects 
of different mass have different clustering properties, 
i.e.  ``correlation lengths'' , 
the latter increasing with object's mass: 
the highest peaks of the density field are more 
 ``strongly clustered''  than the density field itself. 
We have  shown\cite{gsl99,gsld99}  that, in the general case 
of  distributions with a well-defined average  
density, the value at fixed $r$ of $\xi(r)$  is only related 
to the amplitude of the local fluctuation 
with respect to the average density\cite{perezmercader,gsl99,gsld99} 
and it does not give any information of the spatial extension 
of structures in the system. 
Let us see in more detail this point.

The simplest assumption to describe the distribution 
of mass in the universe is that the one of galaxies 
is a good tracer of the distribution of dark matter. 
A specific model has been suggested by Kaiser \cite{kas84} 
in which galaxies and galaxy clusters represent  
different high density peaks of the mass density field. 
Then the term biasing has been used to refer to a number of 
different but related effects \cite{sw94}.  The so-called peaks biasing 
model originally proposed by Kaiser \cite{kas84} 
makes a definitive prediction for the relation  
between the correlation function of galaxies of different 
masses, galaxy clusters (which we generally call {\it objects})  
and dark matter ({\it dm}), at least at large scale: 
\be 
\label{intro1} 
\xi_{obj}(r) = b_{obj}^2 \xi_{dm}(r) \; , 
\ee 
 $b_{obj}$ being the corresponding bias parameter, 
and $\xi_{dm}(r)$ is the correlation function  
of ``dark matter'', i.e. of the underlying density field. 
Rather than  being one bias parameter for 
the correlations of galaxies, there is  an 
undetermined number of such parameters. 
The bias parameter $b_{obj}$ for each class of objects 
is now one of the fundamental  
parameter included both in the theoretical model, 
and in the interpretation of galaxy correlation. 
For instance, for what concerns the clustering of galaxies 
of different luminosity (mass) \cite{par94,ben96}  
 the biasing  is usually referred to 
as {\it luminosity bias}, while for the case of galaxy cluster 
it has been introduced in the {\it clustering-richness relation} 
\cite{bs83}. 
Moreover the ``bias parameter'' plays a crucial role in the 
interpretation of the peculiar velocities  
of galaxies and clusters as well as of the  
anisotropies of the CMBR \cite{sw94}.

The incorrect definition of ``correlation length'' used in  
cosmology \cite{pee80} is not just a question  
of semantics \cite{perezmercader}, but it has generated a  
confusion even when the average density of the system  
is a well-defined property, especially for what concerns the concept of bias 
 \cite{gsld99,gsl99}.  
For instance, we have shown \cite{gsl99} 
 that Eq.\ref{intro1}  increases  the amplitude of $\xi(r)$  
and hence the amplitude of the fluctuations 
with respect to the average density, but 
the typical dimension of structures of  
fluctuations remains the same. 
In order to illustrate  
more clearly this point, let us recall briefly the concept of correlation  
(see Gabrielli \& Sylos Labini\cite{gsl99} for a more detailed discussion). 
If the presence of an object at the point $\vec{r}_1$  
influences the probability of finding another object  
at $\vec{r}_2$,  
these two points are correlated. Hence  there is a correlation 
at the scale distance $r$ if  
\be 
\label{corr1} 
G(r) = \langle n(\vec{0})n(\vec{r})\rangle   \ne \langle n\rangle  ^2   
\ee 
where we average over  
all occupied points of the system  chosen as origin and on the 
total solid angle supposing statistical isotropy. 
On the other hand, there is no correlation if 
\be 
\label{corr2} 
G(r) = \langle n\rangle  ^2. 
\ee 
The proper definition of  $\lambda_0$, the {\it homogeneity scale}, 
is  the length scale beyond which the average density 
becomes to be well-defined, i.e.  
there is a crossover towards homogeneity with a flattening of $G(r)$. 
The length-scale $\lambda_0$ represents the typical dimension 
of the voids in the system. 
On the other hand, the {\it correlation length} $r_c$   
separates correlated regimes  
of the fluctuations with respect to the average density 
from 
uncorrelated ones, 
and it can be defined only if a crossover towards homogeneity is  
shown by the system, i.e. $\lambda_0$ exists \cite{perezmercader}. 
In other words $r_c$ defines the   organization 
in geometrical structures of the fluctuations  
with respect to the average density. Clearly 
$r_c > \lambda_0$: 
only if the average density can be defined  
one may study the correlation length of the fluctuations 
from it. 
In the case in which $\lambda_0$ is finite  
and then $\langle n \rangle >0$, in order  
to study the correlations properties of the fluctuations around the 
average and then the behaviour of $r_c$, we can introduce the  
correlation function $\xi(r)$. 
 
We note that if $\lambda_0 \ll R_s$, $\lambda_0$ 
has nothing to share with questions like  ``which is the typical size  
of structures in the system?'' or  
``up to which length-scale the system is clusterised?'' 
\cite{perezmercader}. 
The answer to this question is strictly related to $r_c$ and not to  
$\lambda_0$.  
The length scale 
 $r_c$   characterizes the  distance over which two different points  
are correlated (clusterised).  
In fact, this property is related not to how large are the  
fluctuations with respect to the average ($\lambda_0$), 
 but to the length extension of their  
correlations ($r_c$).  
 
To be more specific,  
let us consider a fixed set of density fluctuations. 
They can be superimposed to 
different value of a uniform density background. 
The larger is this background the lower $\lambda_0$, but obviously 
the   length scale  
of the correlations ($r_c$) among these fluctuations is not changed, i.e. 
they are clusterised independently of the background (see Fig.\ref{fig1}). 
\bef  
\epsfxsize 8cm 
\centerline{\epsfbox{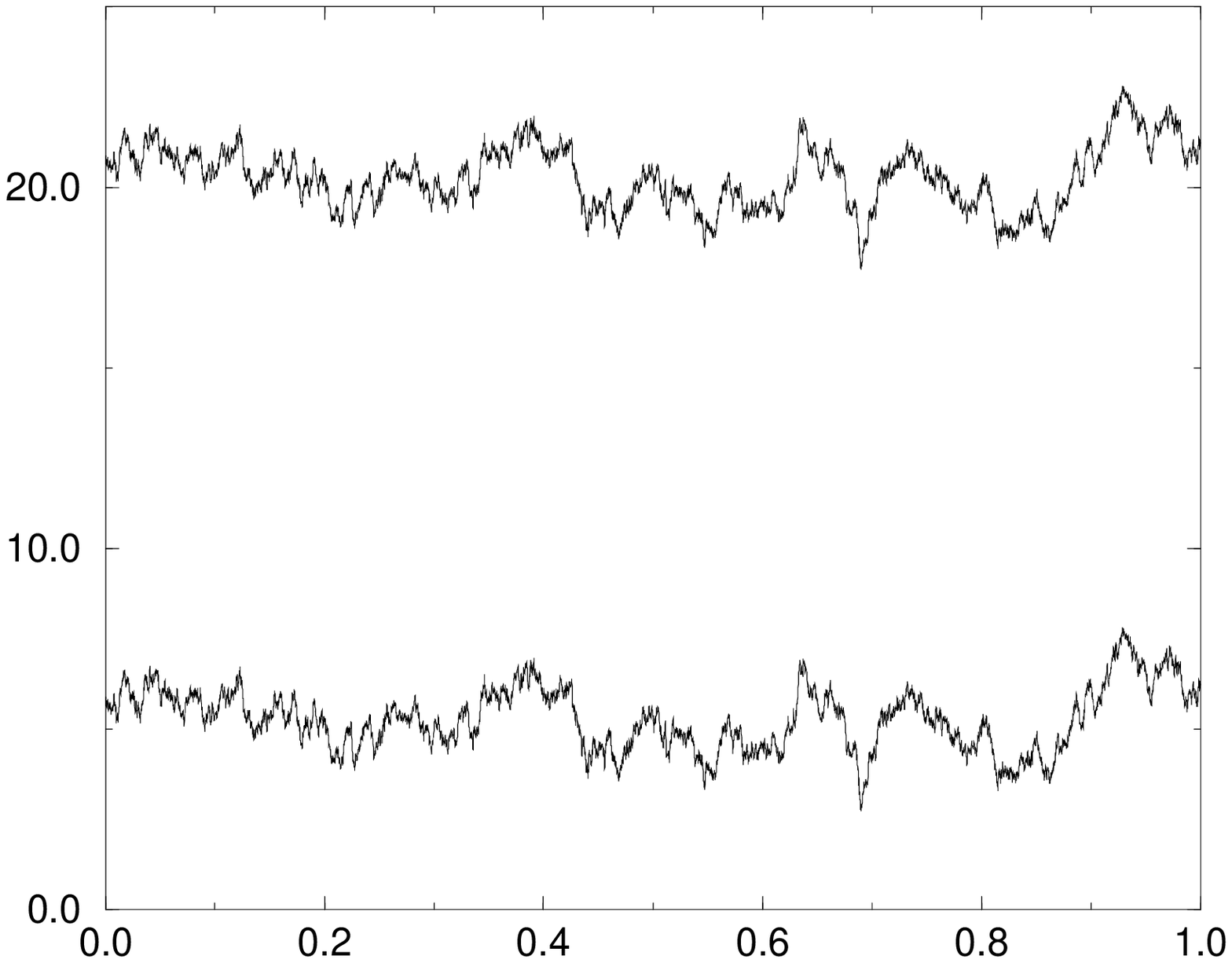}} 
\caption{\label{fig1}  
Guassian  
fluctuations with correlation up to a scale $r_c\approx 0.1$  
in the density field super-imposed on a uniform background.  
The background density (and hence the average density) 
 is  smaller for the lower density field than 
 for the upper one , but  the correlation length is  the 
 same for the two distributions. 
 The amplitude of $\xi(r)$ at the same distance scale,  
is clearly larger 
 for the lower  distribution than for upper one: 
 this is because the amplitude of the fluctuations  
 with respect to the average density is larger. 
 The correlation length $r_c$  is finite and it is related  
 to the largest spatial extension of the  
 fluctuations strctures.  
Beyond $r_c$ the distribution of the fluctuations  
from the average density is completely random.  
} 
\eef 
The conclusion\cite{gsl99} is that a  
 linear amplification of $\xi(r)$ 
\be 
\label{and7} 
\xi'(r)= A \xi(r)  
\ee 
doesn't change $r_c$ (which can be finite or infinite) 
but only $\lambda_0$, i.e. if $A>1$ 
we need  larger subsamples to have a good  
estimation of $\langle n \rangle$, 
but it doesn't change the characteristic length  
(correlation length) of the structures. 
For a more detailed discussion of the concept 
of bias we refer to Gabrielli et al. (1999)\cite{gsld99} 
and Gabrielli \& Sylos Labini(1999)\cite{gsl99}.

\section{Conclusion}

In the discussion about the  
theoretical implication of our results,  
we should not forget the invisible, 
'dark' matter, which is thought to account for at least 90 per 
cent of the mass in the Universe. Apart from the galaxy rotation curves, which is a different evidence, the 
exotic forms of dark matter are introduced to explain the observed puzzling 
properties of visible matter. Actually in the most recent propositions 
there are two weird forms of DM which add to about $98 \%$ of the total matter. 
So the standard interpretation is entirely based on unknown entities whose 
properties are defined just to explain the observed data. 
In our approach we show the correct statistical properties of the visible 
matter which are different than the usual ones. These results in the above 
perspective have important implications for the eventual DM which, however, 
has now to be reconsidered in the new perspective. 
The properties of dark matter 
in the standard picture are inferred from the observed   
properties of visible matter and radiation. 
Now one studies change in these properties  
and in this respect they will have consequences  
on dark matter too\cite{dsl98,gsl99} 
 
For some questions 
the fractal structure leads to a radically new perspective  
and this is hard to accept. But it is based on the best data and  
analyses available. It is neither a conjecture nor a model,  
it is a fact. 
The theoretical problem is that  
there is no dynamical theory to explain 
how such a fractal Universe could have arisen from the pretty 
smooth initial state we know existed in the big bang.  
However this is a  different  
question. The fact that something can be hard to explain theoretically has 
nothing to do with whether it is true or not. Facing a hard problem is far 
more interesting than hiding it under the rug by an inconsistent  procedure.
For example some interesting attempts to understand why 
gravitational clustering generates scale-invariant structures have been
recently proposed by de Vega et al\cite{devega1,devega2,devega3}.
Indeed this will be the key point to understand in the future,  
but first  we should agree on how these new 3d data should be analyzed. 
In addition, the eventual crossover to homogeneity has also to be found with 
our approach. 
If for example homogeneity would really be found say at  
$  \sim100 \hmp$, then clearly 
all our criticism to the previous methods
and results  still holds fully. In summary the 
standard method cannot be used neither to disprove homogeneity, nor to 
prove it. One has simply to change methods.

\section*{Acknowledgements}

We warmly thank   
R. Durrer, A. Gabrielli, 
M. Joyce and M. Montuori  
with whom various parts of these work have 
been done in  fruitful collaborations. 
FSL is grateful to Pedro Ferreira for very useful 
comments and discussions. 
We thank Prof. N. Sanchez and Prof. H. De Vega  
for the organization of this very interesting Conference.  
This work has been partially supported by the  
EC TMR Network  "Fractal structures and   
self-organization"   
\mbox{ERBFMRXCT980183} and by the Swiss NSF.

\end{document}